> REPLACE THIS LINE WITH YOUR PAPER IDENTIFICATION NUMBER (DOUBLE-CLICK HERE TO EDIT) <    1# Why Small, Cold and Quiet DC-DC Conversion is Impossible

Jacques A. Naudé, Ivan W. Hofsajer, *Member, IEEE**Abstract*— Using the first law of thermodynamics and the Parseval-Plancherel theorem, it is shown that every DC-DC converter must trade-off size, heat and conducted electromagnetic noise. It is therefore fundamentally impossible to simultaneously reduce all three of these characteristics to their respective theoretical minimum values.

A figure of merit is introduced which holistically captures the performance of an arbitrary DC-DC converter, this is called the converter's non-ideality and it has a target value of zero. It is derived using the first law of thermodynamics and is shown to be dependent on the efficiency and the root-mean-squared total harmonic distortion of the output voltage.

Interestingly, it is also shown that: boost conversion is impossible without energy storage; ideal rectifiers convert *all* of the input power spectral density into DC (and introduce more noise in the process); the input current of any DC-DC converter scales with the gain of the device squared.

Using an arbitrary DC-DC converter, the culprit of this inherent trade-off is shown to be the act of switching itself. Switching creates harmonics which need to be filtered or transformed into heat in order to get a pure DC voltage at the output of the converter. Even with ideal sub-systems, whether or not resonant conversion is employed, the result stands. The conclusion is that spreading of the switching noise is a primary goal in attempting to reach the impossible.

*Index Terms*—## I. INTRODUCTION

WE do not (at present) possess an AC battery nor a DC transformer and as such, the workhorse of power electronics is the switch. Switching enables the generalised conversion of electrical power from different sources to various loads. As a field, power electronics is considered mature with most of the foundational theoretical work set and improvements to components and packaging/presentation of power conversion services continuing the progress [1]. DC-DC converters are at a point now where their use is anticipated to become ubiquitous as renewable sources of energy enter the mainstream [1]–[3].

Major advancements can be said to have been made on the technological side of power electronics, with modern semi-conductor switching devices approaching the ideal switch [4], [5]. On the theoretical side, the analysis of switched mode power supplies was difficult before the (state-space) averaging and linear ripple approximation created by Middlebrook, Cúk and their research team [6], [7], [8]. The culmination of all of this research resulted in the classic [9].

Definitive answers to basic questions of scale and type are difficult to find. A notable exception is the recent work on scaling laws which govern the behavior of inductors [10]. Questions like, "Is switching fundamentally necessary?", "How big must a device be for a given level of power processing?" and "Is heat an inescapable by-product of conversion, if so, how much is required?" are not available in the introductory and foundational texts [3], [9], [11]. Of interest is the connection between heat and noise. The newer switching technologies have a noted increase in noise even as their switching losses have been reduced [12].

This investigation is a rational argument that every DC-DC converter must operate on the constraint surface implicit in Fig. 1.

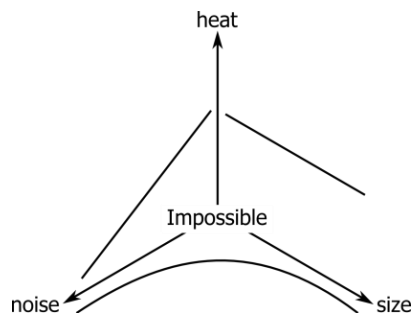

Fig. 1. Salient features of the fundamental constraint surface upon which every DC-DC converter must operate. The ideal case of a small, efficient and quiet converter is theoretically impossible, even with ideal components.

### A. Organisation of Paper

The approach we have taken is to be agnostic about exactly how the DC-DC power conversion process has taken place. Where at all possible, details are removed in favour of a holistic argument involving boundary constraints and known physical laws. Where appropriate, ideal circuit components and processes are defined and reasoned about.

Section II considers a large family of possible time domain signals and analyses the requirements of the average expected power within an arbitrary power converter.

Section III provides a detailed account of DC plus ripple modeling and the implications on the Power Spectral Density

Date submitted:
This work is based on the research supported in part by the National Research Foundation of South Africa (UNIQUE GRANT NO: 98251).
J. A. Naude is a PhD student at the University of the Witwatersrand, Johannesburg, South Africa (e-mail: janaude@gmail.com).
I. W. Hofsajer is head of the Future Electrical Engineering Technology research group at the University of the Witwatersrand, Johannesburg, South Africa (e-mail: ivan.hofsajer@ieee.com).

(PSD) of an arbitrary signal.

Section IV elaborates on the attributes of the PSD of every possible switching function as these are crucial for the line of argument followed throughout the paper.

Sections V-VIII discuss the core thesis of the paper which is that the size of the device, heat dissipated and noise present in any DC-DC converter may not be made arbitrarily small. Appendices are used for the more technical parts of the proofs of each key result.

## II. THE AVERAGE EXPECTED POWER IN A POWER CONVERTER

Given a time varying signal, $f(t)$, the angle bracket of it is defined by

$$\langle f(t) \rangle := \lim_{T \to \infty} \frac{1}{2T} \int_{-T}^{T} f(t) \mathrm{d}t. \tag{1}$$

In addition, it is desireable to include the family of random signals for analysis as well in order to include a broader class of signals and switching schemes. The expected value is denoted by $\mathbb{E}\{\ \}$. Hence the average expected value of $f(t)$ is defined by

$$\langle \mathbb{E}\{f(t)\} \rangle = \lim_{T \to \infty} \frac{1}{2T} \int_{-T}^{T} \mathbb{E}\{f(t)\} \, \mathrm{d}t. \tag{2}$$

with $\mathbb{E}\{f(t)\} := \int \mathbb{P}\{\boldsymbol{\rho}\} f(t) \mathrm{d}\boldsymbol{\rho}$. Note that $\mathbb{P}\{\boldsymbol{\rho}\}$ is the joint probability of $\boldsymbol{\rho}$ which is the set of random variables that characterise the stochastic variation of $f(t)$; the integral in (2) is in the measure theoretic sense [13].

For any machine, the first law of thermodynamics states that energy is locally conserved meaning that it can only be missing from one portion of the machine if it moved to another portion nearby [14], [15]. This is stated in the form of an equation via

$$E_{\mathrm{in}} = E_{\mathrm{out}} + E_{\mathrm{lost}} + E_{\mathrm{stored}}, \tag{3}$$

which is valid at any instant in time. It is physically meaningful to take the time derivative of (3), which will measure the flow of energy into, around and out of the machine. The time derivative is,

$$P_{\mathrm{in}} = P_{\mathrm{out}} + P_{\mathrm{lost}} + P_{\mathrm{stored}}, \tag{4}$$

where all terms should be familiar except for the last term.

The power stored is the time rate of change of the energy stored. If it is positive, the storage components in the machine have a net accumulation of energy and if it is negative then there is a net release. The machines considered for this paper are electrical power converters and hence they do not have a source of power internal to the machine. It is therefore a disallowed condition that the stored power sustains a negative value. It is also a disallowed condition that the stored power sustains a positive value since that would imply that the energy stored will grow to become unbounded. The result would be a fault condition at some point along the trajectory of growth [16].

A consequence of these statements is that, for stable operation of the machine, the average expected storage power must be zero i.e. $\langle \mathbb{E}\{P_{\mathrm{stored}}\} \rangle = 0$ [17]. Hence, taking the average expected value of (4) means that

$$\langle \mathbb{E}\{P_{\mathrm{in}}\} \rangle = \langle \mathbb{E}\{P_{\mathrm{out}}\} \rangle + \langle \mathbb{E}\{P_{\mathrm{lost}}\} \rangle. \tag{5}$$

This statement is a generalized version of the first law of thermodynamics and includes stochastic variations in the power as well. It is universal in the sense that it is valid for any power converter.

The definition of efficiency is given by the relation $\eta := \langle \mathbb{E}\{P_{\mathrm{out}}\} \rangle / \langle \mathbb{E}\{P_{\mathrm{in}}\} \rangle$. Using the fact that the machine cannot supply any energy which it hasn't stored, it is not difficult to prove that $\eta \leq 1$. Using this framework, any average expected power which does not reach the output is considered lost and we use the generic term 'heat' to refer to this lost power. The generic implications of DC plus ripple modeling are elaborated on next.

## III. DC PLUS RIPPLE MODELING

The line of reasoning followed is very similar to conventional small signal analysis for the standard DC plus ripple model used in DC-DC power converter applications [3], [8], [9]. There is one key difference; the additive signal is a (possibly) random process and the ripple is not necessarily small. Let the variable under consideration be denoted by $x(t)$. The DC plus ripple model of this variable is then given by

$$x(t) = X + \tilde{x}(t) \tag{6}$$

where

$$X := \langle \mathbb{E}\{x(t)\} \rangle \tag{7}$$

and hence,

$$\langle \mathbb{E}\{\tilde{x}(t)\} \rangle = 0. \tag{8}$$

The proof of equation (8) is by definition. Consider taking the average expected value of equation (6), the result is $\langle \mathbb{E}\{x(t)\} \rangle = X + \langle \mathbb{E}\{\tilde{x}(t)\} \rangle$ which implies that $\langle \mathbb{E}\{\tilde{x}(t)\} \rangle = \langle \mathbb{E}\{x(t)\} \rangle - X$. But by definition (7) this must be $\langle \mathbb{E}\{\tilde{x}(t)\} \rangle = X - X = 0$.

Lastly, it can be shown that

$$\left\langle \mathbb{E}\left\{\frac{\mathrm{d}\tilde{x}}{\mathrm{d}t}\right\} \right\rangle = 0. \tag{9}$$

Derivatives make additive constants zero and amplify noise, even for random processes [18]. Hence the time derivative of either a random process or a deterministic function of time which has an average expected value of zero.

## IV. THE PARSEVEL-PLANCHEREL EQUATION

The Parseval-Plancherel theorem is an important result [18]. It states that,

$$\int_{-\infty}^{\infty} \mathcal{S}_{xx}(f) \, \mathrm{d}f = \langle \mathbb{E}\{x(t)^2\} \rangle, \tag{10}$$

and

$$\int_{-\infty}^{\infty} \mathcal{S}_{xy}(f) \mathrm{d}f = \langle \mathbb{E}\{x(t)y(t)\} \rangle = \int_{-\infty}^{\infty} \mathcal{S}_{yx}(f) \mathrm{d}f. \tag{11}$$

Note that $\mathcal{S}_{xx}(f)$ is the PSD of $x(t)$ and $\mathcal{S}_{xy}(f)$ is the cross-PSD of $x(t)$ and $y(t)$. This theorem is crucial for the calculation of actual power in the exposition to follow.

*A. PSD of DC plus Ripple Equation*

The following is universally true about the PSD for both random and deterministic signals which can be modelled by the DC plus ripple condition in (6),

$$S_{xx}(f) = S_{\widetilde{x}\widetilde{x}}(f) + X^2\delta(f). \qquad (12)$$

The proof of this assertion is in the appendix. In the case of DC-DC conversion, it makes sense to call $S_{\widetilde{x}\widetilde{x}}(f)$ the noise PSD of $x(t)$, since anything which is not DC is considered to be noise. This result is useful for separating out the DC from the noise of the output voltage in an arbitrary DC-DC converter.

*B. PSD of Every Possible Switching Function*

Consider a switching function $q(t)$ which has the properties that:
1. The switching function can either be 1 or 0 at any instant in time.
2. The complement of the switching function is denoted by $q'(t) \coloneqq 1 - q(t)$.
3. The time derivative of the switching is not well-defined and is therefore represented using weighted Dirac delta functions at the transition times.

The switch has an average expected value of $\langle\mathbb{E}\{q(t)\}\rangle = D$ which implies that $\langle\mathbb{E}\{q'(t)\}\rangle = (1-D) = D'$. The value $D$ is typically called the duty-cycle in PWM but is far more general in this context. It would, for example, include the average expected value of any type of Random PWM scheme as well.

Using this definition, the total PSD of every possible switching function can be shown to be $\langle\mathbb{E}\{q(t)^2\}\rangle = D$. Hence, using the DC plus ripple model, the PSD of every possible switching function can be written as

$$S_{qq}(f) = DD'g(f) + D^2\delta(f) \qquad (13)$$

where $g(f)$ represents the distribution of the switching noise in the frequency domain. It has the following important properties.

**Property 1**: $g(f) \geq 0$ for all frequencies
**Property 2**: $g(f) = g(-f)$
**Property 3**: $\int_{-\infty}^{\infty} g(f)\mathrm{d}f = 1$

These properties can be derived using the Parseval-Plancherel constraint, the DC plus ripple model in (12), the positive definite property of the PSD of any real signal and the average expected requirement $\langle\mathbb{E}\{q(t)\}\rangle = D$.

Note that the total switching noise is $DD' = D(1-D)$ for any possible switching function. What would be different amongst the various possible switching schemes is the distribution of the noise, which is described by $g(f)$.

V. A SWITCHING RESISTIVE LOADED DC-DC CONVERTER

Consider now the resistive loaded DC-DC converter depicted in Fig. 2, the internals are arbitrarily driven by a switching function, defined previously as $q(t)$.

For this case it is considered that the DC-DC converter has two circuit configurations only. The generalization to include more configurations is a modification of the present exposition.

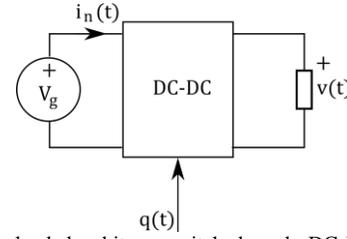

Fig. 2. Resistive loaded, arbitrary switched mode DC-DC converter. The switching function is shown to enter the control port of the circuit as it is conventionally depicted. The switching function alters the arrangement of passive components only.

The input voltage $V_g$ is considered to be ideal and therefore it has zero ripple. The output power is calculated in terms of the voltage; the DC plus ripple model is applied to all of the time domain variables, Parseval-Plancherel's theorem is used and the results are incorporated into the extended version of the first law of thermodynamics in (5).

The result is that

$$\begin{aligned}
V_g I_n &= \langle\mathbb{E}\{P_{\text{loss}}\}\rangle + \frac{1}{R}(V^2 + \langle\mathbb{E}\{\tilde{v}^2\}\rangle) \\
V_g I_n - \frac{V^2}{R} &= \langle\mathbb{E}\{P_{\text{loss}}\}\rangle + \int_{-\infty}^{\infty}\frac{1}{R}S_{\widetilde{v}\widetilde{v}}(f)\mathrm{d}f
\end{aligned} \qquad (14)$$

where in the final step the average output voltage squared has been moved over to the left hand side. By definition of DC-DC conversion $V = GV_g$, and hence

$$V_g I_n - \frac{G^2 V_g^2}{R} = \langle\mathbb{E}\{P_{\text{loss}}\}\rangle + \int_{-\infty}^{\infty}\frac{1}{R}S_{\widetilde{v}\widetilde{v}}(f)\mathrm{d}f \qquad (15)$$

Our main thesis is presented in (15). In plain English it states that *something = heat + total noise power*. We choose to call this *something* the non-ideal power, $n_I$ and a new figure of merit may defined in terms of it.

*A. Non-Ideality: A New Figure of Merit*

The non-ideality, $\epsilon_I$, is defined by

$$\epsilon_I \coloneqq \frac{n_I}{\langle\mathbb{E}\{P_{in}\}\rangle} = (1-\eta) + \eta\,\text{VTHD}_R^2, \qquad (16)$$

and it measures the fraction of input power which is in the incorrect form i.e. heat and/or noise.

The non-ideality is related to the efficiency and the RMS total harmonic distortion of the output voltage (VTHD$_R$). The derivation of (16) from (15) is accomplished in the appendix. Note that $0 \leq \text{VTHD}_R^2 \leq 1$.

The target goal of the non-ideality is zero and it takes into account losses of all kinds as well as the conducted noise on the output voltage. The closer the non-ideality is to zero, the closer any given DC-DC converter is to ideal. Note that it is topologically agnostic and is independent of how many switching functions are required to describe the DC-DC converter since it was derived from the first law of thermodynamics and Parseval-Plancherel's theorem. It is a relationship that is true by definition. Contours of this function are depicted in Fig. 3.

An important note is that VTHD$_R$ is a function of the size of the device, it may be made arbitrarily small (theoretically)

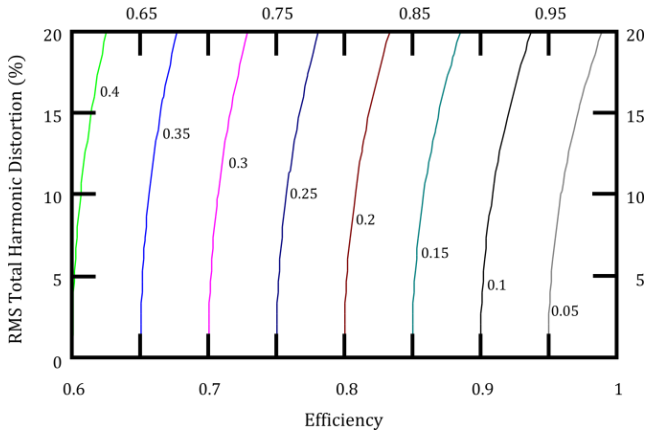

Fig. 3. Contour lines of constant non-ideality, $\epsilon_I$, plotted against efficiency, $\eta$, and RMS total harmonic distortion, $\text{VTHD}_R$. Observe that below approximately 5% total harmonic distortion, the non-ideality and efficiency are approximately equivalent.

through the use of energy storage components which filter the output voltage. Hence, larger filtering components mean lower total harmonic distortion. This notion will be made precise in the sequel.

### B. Fundamental Input Current Requirement

As a brief but interesting aside; (15) also produces a scaling law for the average input current in any resistive loaded DC-DC converter. Since total output voltage noise PSD must be positive or zero, the implication is that

$$\frac{G^2}{\eta}\left(\frac{V_g}{R}\right) \leq I_n. \qquad (17)$$

This is true for *any* DC-DC converter and it was derived using the first law of thermodynamics, the definition of power loss and power spectral density. It is a hence a very strong inequality. The proof is in the appendix.

There are a number of notable features of (17). If $q(t)$ simply connected the supply to the load, $V_g/R$ is the current which would flow. As a result of DC-DC conversion, a different input current is required. The additional current required is proportional to the square of the gain and inversely proportional to the efficiency of the device. Real converters will include source impedances, conductive losses due to resistance in non-super conducting elements etc. and these will have a bearing on the amount of input current needed (through the efficiency). Hence, (17) is the theoretical minimum input current requirement for any DC-DC converter.

We now turn to our main method of proof of the impossibility of the ideal DC-DC converter. We make two aspects of the converter a minimum and prove that the third cannot be made arbitrarily small as well. This is accomplished for each of the three possibilities.

## VI. SMALL AND COLD BUT NOISY

Without energy storage components or swapping polarity of the source, boost conversion is fundamentally impossible. The proof is in the appendix.

Hence, consider a standard buck converter with ideal switches and no filtering components. Let the conduction losses be zero as well so that the only resistive component in the buck converter is the load i.e. $\eta = 1$.

Both size and heat are at their minimum values in this example; this is the smallest DC-DC converter possible (dependent only the physical size of the switching components). Since it has an efficiency of unity, temperature effects within the converter do not exist and hence it is considered to be 'cold'.

Using the fact that $q(t) = D + \tilde{q}(t)$, the output voltage is equal to

$$v(t) = q(t)V_g = DV_g + \tilde{q}(t)V_g. \qquad (18)$$

Taking the expected average of the output voltage leads to $\langle \mathbb{E}\{v(t)\}\rangle = DV_g$ as one would expect from a standard buck converter. The output voltage has the correct expected average but the shape is from the square-wave family, since there are no filtering components. This is represented by $\tilde{q}(t)$ which is the switch deviation from the average.

Note, the efficiency is unity, $\eta = 1$, the DC output voltage is correct but the output voltage is extremely noisy. The non-ideality, using (12) and (18), is equal to

$$\epsilon_I = \text{VTHD}_R^2 = \left(\frac{\int_{-\infty}^{\infty} V_g^2 S_{\tilde{q}\tilde{q}}(f)\mathrm{d}f}{D^2 V_g^2 + \int_{-\infty}^{\infty} V_g^2 S_{\tilde{q}\tilde{q}}(f)\mathrm{d}f}\right) = \frac{DD'}{D}. \qquad (19)$$

Given that $D$ is used to specify the average output voltage, there are no degrees of freedom left with which to alter the non-ideality. The non-ideality is completely composed of noise even though the DC-DC converter has the correct average output voltage and the efficiency is unity. The conclusion is that a small, cold buck converter is very noisy. In fact the noise is the worst possible, since, for a given total power spectrum of noise, the square-wave family signal is extremal [19], [20]. The only benefit this kind of converter has is that the volume is made up entirely of the volume of the switching elements; and is therefore a minimum.

Resonant DC-DC conversion is out of the question since it would require a transformer (in general) in order to achieve gain. Transformers (like the filtering components) take up space and therefore the volume is no longer minimized. Resonant DC-DC converters have other fundamental problems which will be addressed in the sequel.

## VII. QUIET AND COLD BUT LARGE

Consider now an arbitrary resistive loaded DC-DC converter. Assume again that conduction losses are zero and the only conducted EMI filtering that is done is lossless i.e. there are no equivalent resistors anywhere within the converter. Hence, $\eta = 1$ and therefore $\epsilon_I = \text{VTHD}_R^2$.

There are two cases possible now, non-resonant DC-DC conversion and resonant DC-DC conversion.

### A. Non-Resonant DC-DC Conversion

It is possible to define a switch to output voltage transfer function using the standard framework [8]. The transfer function due to duty cycle variation in Ćuk's model is a good approximation for the switching ripple to output voltage transfer function, $H_{\widetilde{qv}}$. This is proved in the appendix. In addition, it is near identical to the best linear approximation defined in [21].

Using the standard technique from a linear systems theory with random inputs, the non-ideality will, in general, be equal to

$$\epsilon_I = \text{VTHD}_R^2 = \frac{DD' \int_{-\infty}^{\infty} |H_{\widetilde{qv}}|^2 g(f) df}{G^2 V_g^2 + DD' \int_{-\infty}^{\infty} |H_{\widetilde{qv}}|^2 g(f) df} \quad (20)$$

where $G$ is the DC voltage gain and $g(f)$ is the distribution of the switching noise.

Note the effect of the filter is represented by $H_{\widetilde{qv}}$, this quantifies how much of the switching noise, $g(f)$, will proceed through into the voltage noise. No conduction losses are permitted (since these would break the condition that $\eta = 1$) and therefore the magnitude of $H_{\widetilde{qv}}$ will be entirely dependent on the size of the inductors and capacitors present, as well as on the circuit topology.

Since the device is a DC-DC converter, $H_{\widetilde{qv}}$ will be of a low-pass type with damping provided by the load resistor since no resistors are present anywhere else in the circuit. Hence, $H_{\widetilde{qv}}(f) \approx \theta(f + f_B) - \theta(f - f_B)$ where $\theta(\cdot)$ is the Heaviside step function and $f_B$ is the bandwidth of the low pass filter. The bandwidth, $f_B$, will be proportional to either $(LC)^{-1/2}$, $RL^{-1}$ or $(RC)^{-1}$. Note these are the main filtering inductor and/or capacitor. In addition, space is needed to house the inductor and/or capacitor with a larger value taking up a larger volume. Hence, $f_B \propto (\text{Vol})^{-1}$ and therefore

$$\int_{-\infty}^{\infty} |H_{\widetilde{qv}}|^2 g(f) df \approx \int_{-f_B}^{f_B} g(f) df \propto \text{Vol}^{-1}, \quad (21)$$

because $0 \le g(f) < 1$ using Property 1,2 and 3. The final result, using (20) and (21) means that the non-ideality will be proportional to

$$\epsilon_I \propto \frac{DD'}{\text{Vol } G^2 V_g^2 + DD'}. \quad (22)$$

Since $0 < DD' < 1, G \neq 0, V_g \neq 0$ for a functioning DC-DC converter, the non-ideality of any non-resonant DC-DC converter is inversely proportional to the volume of its main filtering elements. The non-ideality may therefore be made smaller only at the cost of increasing the volume of the device, even with ideal components.

Hence, a cold, quiet non-resonant DC-DC converter must necessarily be large.

### B. Resonant DC-DC Conversion

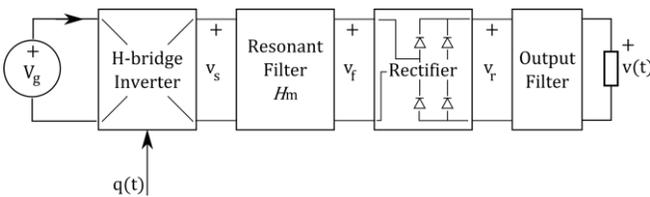

Fig. 4. A generic block diagram for a resonant DC-DC converter with an uncontrolled rectifier at the output stage. The H-bridge inverter and rectifier are considered to be ideal and have an efficiency of unity for the purpose of the argument.

A generic resonant DC-DC converter with controlled H-bridge inverter as input source and uncontrolled rectifier at the output is depicted in Fig. 3.

For the sake of the argument, both the inverter and rectifier have an efficiency of unity and the resonant filter has a voltage transfer function denoted by $H_m(j\omega)$. The line voltage $V_g$ undergoes inversion to become $v_s(t)$ which is filtered by $H_m$ to become $v_f$, where $f$ is a mnemonic for *filtered* source voltage. This filtered voltage is rectified to become $v_r$ which is filtered by the output filter before being dropped across the load, where it is considered to be the output voltage $v(t)$.

*1) The H-Bridge Inverter Creates Switching Harmonics*

By swapping polarity with the switching dictated by $q(t)$, the source voltage is equal to

$$v_s(t) = q(t)V_g - q'(t)V_g = (2D - 1)V_g + 2\tilde{q}V_g \quad (23)$$

where the DC plus ripple model was applied to both $q$ and $q'$ to arrive at the final form. To simplify the analysis, it is assumed that $q(t)$ is a PWM waveform with a 50% duty cycle that is switched at a switching frequency $f_s = 1/T$. Under this condition, the source voltage is equal to $v_s(t) = 2\tilde{q}V_g$ where the peak and trough values of $\tilde{q}$ are $\pm 1/2$. The total PSD of the switching noise is $DD' = 1/4$.

The PSD of the source voltage is given by $S_{v_s v_s}(f) = V_g^2 g(f)$ whereas the PSD of the un-switched line voltage was $S_{V_g V_g}(f) = V_g^2 \delta(f)$.

Note how the line voltage PSD evolves from a DC Dirac delta to become a distribution in the frequency domain with $g(f)$ after the inverter function. It can therefore be said that the inverter has created harmonics $g(f)$ which were not present before.

*2) The Resonant Filtered PSD*

The usual describing function analysis has been foregone in this analysis [3]. We have done this in order to keep track of what happens to the "noise" spectrum i.e. the unwanted oscillations. The resonant filter, when applied to the source voltage has the result that

$$S_{v_f v_f}(f) = |H_m|^2 S_{v_s v_s}(f) = V_g^2 |H_m|^2 g(f). \quad (24)$$

The usual assumption is that the output of the resonant filter is dominated by the sinusoid which results from the resonant driving signal, $v_s$ [3]. In our case, we do not make this assumption, we instead include all of the noise present in $g(f)$ in (24). This is also known as the best linear approximation [21], [22].

*3) The Rectifier Introduces More Noise*

It has been possible to use linear techniques in calculating the filter voltage due to the filtering of random inputs theorem [18], [23]. It is not possible to define a transfer function for the ideal rectifier, hence the reason for the usual describing function approach. Stated another way, it is not possible to calculate the PSD of the rectified voltage using linear techniques.

The ideal rectifier output voltage is denoted by $v_r(t) = |v_f(t)|$ where the absolute value is the hard non-linearity which prevents linear circuit theory from being used. Ignoring this difficulty for now, consider calculating the power spectral density of the rectified voltage (if the time domain waveform, including noise, were known analytically).

The PSD would be calculated by the Wiener-Khintchine theorem with $S_{v_r v_r}(f) = \mathcal{F}\{\langle \mathbb{E}\{v_r(t)v_r(t+\tau)\}\rangle\}$. Substituting in the non-linearity means that

$$S_{v_r v_r}(f) = \mathcal{F}\left\{\left\langle \mathbb{E}\left\{\sqrt{v_f(t)^2 v_f(t+\tau)^2}\right\}\right\rangle\right\} \quad (25)$$

where $|x| = \sqrt{x^2}$ has been used as an equivalent hard non-linearity.

A non-trivial result is discovered at $\tau = 0$. The implication, by taking the inverse Fourier transform is that the auto-correlation function of the rectified voltage at zero lag is equal to the *total* PSD of the resonant filtered output voltage i.e. $\mathcal{R}_{v_r v_r}(0) = \langle \mathbb{E}\{v_f^2\}\rangle = \int_{-\infty}^{\infty} S_{v_f v_f}(f) \mathrm{d}f$.

Since $\mathcal{R}_{v_r v_r}(\tau)$ is *not* a Dirac delta, weighted by this amount, the implication of this is that the ideal rectifier has created more harmonics. This is because all of the filtered voltage harmonics are already accounted for at zero lag.

*4) The Ideal Rectifier Perfectly Converts Noise into DC*

Consider now the DC output voltage,

$$\langle \mathbb{E}\{v\}\rangle = \langle \mathbb{E}\{v_r\}\rangle = \langle \mathbb{E}\{|v_f|\}\rangle = \left\langle \mathbb{E}\left\{\sqrt{v_f^2}\right\}\right\rangle. \quad (26)$$

It is assumed that the output filter, being passive, has a gain of unity at DC, otherwise it would be lossy. This condition allows for the assertion that the DC output voltage is equal to the DC rectified voltage, $\langle \mathbb{E}\{v\}\rangle = \langle \mathbb{E}\{v_r\}\rangle$. The absolute value is transformed into an equivalent non-linearity with the square-root of the square of the voltage waveform in the last part of the equality in (26).

The reason for going through all of this analysis is because the DC plus ripple model of the filtered output voltage may now be applied in the argument of (26). Using the fact that the average voltage in the resonant part of the circuit is zero i.e. $V_f = \langle \mathbb{E}\{v_f\}\rangle = 0$; the result is that

$$\langle \mathbb{E}\{v\}\rangle = \left\langle \mathbb{E}\left\{\sqrt{\tilde{v}_f^2}\right\}\right\rangle \approx \mathrm{RMS}_{\widetilde{v_f}} = \sqrt{\int_{-\infty}^{\infty} S_{\widetilde{v_f}\widetilde{v_f}}(f)\mathrm{d}f}. \quad (27)$$

This approximation has much intuitive appeal. It states that the average output voltage applied across the load is approximately equal to the root-mean-squared (RMS) value of the resonant filtered voltage. The mathematical justification of this approximation uses Jensen's inequality, whereby $\langle \mathbb{E}\{u^{1/2}\}\rangle \geq \langle \mathbb{E}\{u\}\rangle^{1/2}$ provided that $u \geq 0$ [24].

What (27) states is that the load DC voltage is lower bounded by the total frequency response of the resonant filtered voltage, $v_f$ i.e. *all* of the switching noise and resonant response of the resonant filter is converted into perfect DC output voltage.

This fact assists with the accounting process of noise in resonant DC-DC conversion. In addition, it makes an interesting prediction. Using the resonant DC-DC converter topology in Fig. 3, it is possible to use any kind of switching scheme which has a duty cycle of 50% with any kind of filtering and, provided only that the RMS voltages are the same, the DC output voltage should be identical. The only assumption is that the rectifier is near ideal, which is a good assumption given current technological progress [25].

Fig. 5 demonstrates the previous exposition pictorially. The resonant filter and output filter are band-pass filters which have a similar integral constraint that shows that the volume constraint cannot be avoided, even when using ideal components.

## VIII. QUIET AND SMALL BUT HOT

Consider again zero conductive losses. By transforming the switching noise into heat, the volume constraint may be relaxed. This would necessarily result in a non-zero non-ideality due to the inefficiency brought about by converting the noise PSD into heat. This requires a filter which captures all non-DC harmonics and converts them entirely into heat.

With very high frequency switching, assuming ideal switching elements, the noise PSD is spread into higher harmonics which require smaller inductors and/or capacitors and therefore result in a smaller device due to the reduced filtering burden. The device is quiet and small but has a minimum amount of heat required due to transforming the entire noise PSD into heat.

The non-ideality becomes equal to (28) in this case.

$$\epsilon_I = (1 - \eta) = \frac{\int_{-\infty}^{\infty} \frac{1}{R} S_{\widetilde{v}\widetilde{v}}(f)\mathrm{d}f}{\langle \mathbb{E}\{P_{\mathrm{in}}\}\rangle} \quad (28)$$

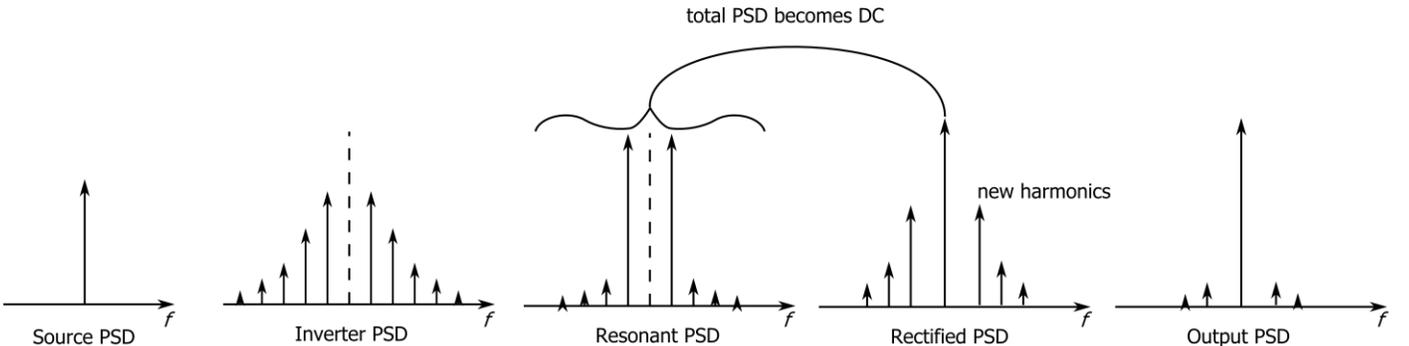

Fig. 5. Evolution of the PSD of the DC voltage source feeding a generic resonant DC-DC converter. Linear techniques are used up to the filtered PSD. A non-linear argument shows that the DC value of the rectified PSD is equal to the total PSD of the filtered PSD. In addition, the rectified PSD introduces new harmonics which need to be filtered at the output stage. The output PSD therefore has a DC value equal to the total PSD of the filtered PSD plus filtered harmonics from rectification.

## IX. Spread of Noise

The volumetric constraint can be asymptotically reduced through modification of $g(f)$, the switching noise. The integral $\int_{-f_B}^{f_B} g(f) \, df$ can only be reduced by ensuring that $g(f) \approx 0$ within the bandwidth of the filter. This can be achieved by either switching at higher frequencies or by using switching schemes which spread the noise, as would be the case with RPWM and related schemes [26].

## X. Conclusion

Heat, noise and volume are inextricably linked in every DC-DC converter. It is fundamentally impossible to reduce all three down to an arbitrarily small value. As components approach the ideal, it is anticipated that this trade-off will become ever more important.

DC-DC converters can now be ranked in a unified and consistent manner. Any DC-DC converter which has a lower non-ideality for a given volume is to be preferred over another. Measuring the non-ideality of a given DC-DC converter is straight forward; $V^2/R$ must be subtracted from the average input power and this difference must be divided by the average input power. The closer the non-ideality is to zero, the better. The reader should find that the heat and noise trade-off will be found to be ubiquitous in their work and in the literature.

## Appendix

It will be proven that transients of $f(t)$ will not contribute to the value of the angle bracket and furthermore that periodic parts will be averaged out with additive constants left unchanged.

### A. Proof: Transients do not contribute

Except for random components, in power electronics it is always possible to split a signal $f(t) = f_\delta + f_p + \alpha$ where $f_\delta$ is the transient part of the signal, $f_p$ is the periodic part of the signal and $\alpha$ is the DC offset. The transient part of the signal has the property that it has a finite "energy", which is to say that it is square-integrable [27].

Integration is a linear operator, hence

$$\langle f(t) \rangle = \langle f_\delta \rangle + \langle f_p \rangle + \langle \alpha \rangle. \tag{29}$$

The value of the integral of $f_\delta$ will tend to a constant since it is square-integrable. Hence, the angle bracket operator on $f_\delta$ results in

$$\langle f_\delta \rangle = \lim_{T \to \infty} \frac{1}{2T} \times \text{const} \to 0. \tag{30}$$

### B. Proof: Periodic parts average

The periodic part of the signal $f_p$, has the property that $f_p(t + \tau) = f_p(t)$ where $\tau$ is the periodicity of the signal. Now the angle bracket integral can be calculated as follows,

$$\langle f_p \rangle = \lim_{k \to \infty} \frac{1}{(2k+1)\tau} \sum_{m=-k}^{k} \int_{m\tau}^{(m+1)\tau} f_p \, dt \tag{31}$$

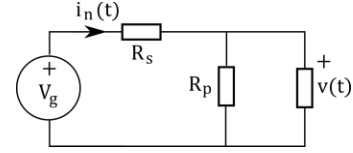

Fig. I. Equivalent circuit of every non-polarity reversing, possibly switched DC-DC converter without any energy storage elements.

where the integral has been broken up into $2k + 1$ consecutive periods and the time variable $T$ is redefined in terms of these. It is assumed that $f_p$ is not some pathological function which does not permit this changing of the limits since only physically realisable functions which represent voltages and currents are considered. By definition of periodicity, the integral over one period will result in the same value as the integral over any other period, hence

$$\sum_{m=-k}^{k} \int_{m\tau}^{(m+1)\tau} f_p \, dt = \sum_{m=-k}^{k} \text{const} \tag{32}$$
$$= (2k + 1) \times \text{const}$$

which implies that the angle bracket of a periodic function will be

$$\langle f_p \rangle = \frac{1}{\tau} \times \text{const} \tag{33}$$

since the $k$'s cancel. Using the result that the constant in equation (33) is the integral over any single period, the angle bracket for a periodic function is exactly equal to

$$\langle f_p \rangle = \frac{1}{\tau} \int_0^\tau f_p \, dt \tag{34}$$

where $\tau$ is the period. This is exactly the time average value of $f_p$.

### C. Proof: Additive constants are left unchanged

With $f(t) = \alpha$ the angle bracket integrates to

$$\langle \alpha \rangle = \lim_{T \to \infty} \alpha \frac{T - (-T)}{2T} = \alpha. \tag{35}$$

It has been shown that the angle bracket operator is linear, ignores transients, averages periodic functions and leaves additive constants unchanged. Taking the angle bracket of a vector signal is straight forward since the dimension of the vector preserved by the integral, each row of the vector is time integrated separately.

### D. Proof: DC plus ripple PSD representation

The proof of (12) uses the Wiener-Khintchine theorem which relates the PSD to the Fourier transform of the auto-correlation function [18]. Explicitly this is given by,

$$S_{xx}(f) \coloneqq \mathcal{F}\{\langle \mathbb{E}\{x(t)x(t+\tau)\}\rangle\} = \mathcal{F}\{\langle \mathbb{E}\{xx''\}\rangle\}, \tag{36}$$

where to simplify the notation, the formal replacement $x(t) \to x$ and $x(t + \tau) \to x''$ was used [18].

Apply the DC plus ripple model to the right hand side of (36) using the formal notation,

$$\begin{aligned}
&\langle \mathbb{E}\{(X + \tilde{x})(X + \tilde{x}'')\}\rangle \\
&= \langle \mathbb{E}\{X^2 + \tilde{x}X + X\tilde{x}'' + \tilde{x}\tilde{x}''\}\rangle \\
&= X^2 + X\langle \mathbb{E}\{\tilde{x}\}\rangle + X\langle \mathbb{E}\{\tilde{x}''\}\rangle + \langle \mathbb{E}\{\tilde{x}\tilde{x}''\}\rangle \\
&= \langle \mathbb{E}\{\tilde{x}\tilde{x}''\}\rangle + X^2,
\end{aligned} \tag{37}$$

which used linearity of the expectation and bracket operator for the third step, and the definition of the average expected ripple being zero in the last step. The proof of (12) requires taking the Fourier transform of (37) and using the definition of the PSD in (36).

*E. Proof: Minimum average input current*

Proving (17) uses (15), $\langle\mathbb{E}\{P_{\text{in}}\}\rangle - \langle\mathbb{E}\{P_{\text{loss}}\}\rangle = \eta\langle\mathbb{E}\{P_{\text{in}}\}\rangle$, and the fact that the total noise PSD $\geq 0$. Explicitly, using (15), the equality is given by,

$$\langle\mathbb{E}\{P_{\text{in}}\}\rangle - \langle\mathbb{E}\{P_{\text{loss}}\}\rangle - \frac{G^2 V_g^2}{R} = \int_{-\infty}^{\infty} \mathcal{S}_{\widetilde{v}\widetilde{v}}(f)\mathrm{d}f$$
$$\Rightarrow \eta\langle\mathbb{E}\{P_{\text{in}}\}\rangle - \frac{G^2 V_g^2}{R} = \int_{-\infty}^{\infty} \mathcal{S}_{\widetilde{v}\widetilde{v}}(f)\mathrm{d}f. \quad (38)$$

All that remains to complete the proof is to use the definition of input power, the fact that the total noise PSD $\geq 0$ and solve for the input current.

*F. Proof: Boost Conversion Impossible without Storage*

Without energy storage, only switches and resistors may be employed. It can be shown, using equivalent resistors that every possible circuit without energy storage is a (possibly switched) version of the circuit depicted in Fig. I.

Let $R_p \to \infty$ and therefore the output current is equal to the input current at all times, $i_n(t) = i_o(t)$. By definition, the efficiency is calculated by $\eta\langle\mathbb{E}\{i_n\}\rangle V_g = \langle\mathbb{E}\{i_o^2 R\}\rangle$, hence

$$\eta V_g = \frac{I_o^2 R}{I_o} + \frac{\langle\mathbb{E}\{\widetilde{i_o}^2\}\rangle R}{I_o} = V + \frac{\langle\mathbb{E}\{\widetilde{i_o}^2\}\rangle R}{I_o}. \quad (39)$$

Re-arranging (39) by factorizing the right-hand side, using Ohm's law and then solving for the average output voltage leads to

$$V = \eta V_g \left(\frac{I_o^2}{I_o^2 + \langle\mathbb{E}\{\widetilde{i_o}^2\}\rangle}\right)$$
$$= \eta V_g (1 - \text{ITHD}_R^2), \quad (40)$$

where $\text{ITHD}_R^2$ is the output current RMS total harmonic distortion (considering the DC value to be the fundamental). Observe that heat and noise influence the average output voltage.

With no switching, there is no noise in the output current, $\text{ITHD}_R^2 = 0$ and therefore $\eta V_g = V$. In other words, the gain is equal to the efficiency, $G = \eta$. The efficiency cannot be over unity and therefore neither can the gain. Boost conversion is impossible in this case.

When switching is present, $R_s$ may be used to implement it. Ideal switching would have $R_s$ alternate between $0$ and $\infty$, non-ideal switching would have finite values in both states. Under switching action $\text{ITHD}_R^2 \neq 0$ and since $0 \leq \text{ITHD}_R^2 \leq 1$ and $0 \leq \eta \leq 1$, the result is that $V \leq V_g$ since the product of two numbers less than 1 is also less than 1.

Inclusion of $R_p$ or finite values of $R_s$ only worsens these results since these conditions dissipate power. Note that (40) has $i_n(t) = i_o(t)$ as a main premise. Hence if $R_p$ is a finite value, the simple relationship in (40) is no longer valid, it would need to be multiplied by a term using the current divider rule. The overall result remains the same.

The above analysis exhausts all possibilities of switched and non-switched DC-DC power converters with no energy storage. Boost conversion is therefore fundamentally impossible without energy storage.

*G. Proof: Total harmonic distortion relationship*

The result is most easily secured by considering the DC value to be the "fundamental" from the definition of the RMS total harmonic distortion. Using the result that $\langle\mathbb{E}\{P_{in}\}\rangle = \langle\mathbb{E}\{P_{out}\}\rangle/\eta$ the total noise integral can be represented as

$$\int_{-\infty}^{\infty} \frac{1}{R\langle\mathbb{E}\{P_{in}\}\rangle}\mathcal{S}_{\widetilde{v}\widetilde{v}}(f)\mathrm{d}f = \int_{-\infty}^{\infty} \frac{\eta}{\langle\mathbb{E}\{v^2\}\rangle}\mathcal{S}_{\widetilde{v}\widetilde{v}}(f)\mathrm{d}f \quad (41)$$

which is the definition of RMS total harmonic distortion using the fact that $\langle\mathbb{E}\{v^2\}\rangle = V^2 + \langle\mathbb{E}\{\widetilde{v}^2\}\rangle$ and considering the DC voltage, $V$, to be the "fundamental" harmonic i.e.

$$\int_{-\infty}^{\infty} \frac{\eta}{\langle\mathbb{E}\{v^2\}\rangle}\mathcal{S}_{\widetilde{v}\widetilde{v}}(f)\mathrm{d}f = \eta \frac{\langle\mathbb{E}\{\widetilde{v}^2\}\rangle}{V^2 + \langle\mathbb{E}\{\widetilde{v}^2\}\rangle}. \quad (42)$$